\documentclass[1p]{elsarticle}

\usepackage{lineno,hyperref}

\modulolinenumbers[2]

\journal{NIM-A}

\newcommand{\eemmtt}{\ensuremath{e^+ e^- \to \mu^+ \mu^- \tau^+ \tau^-}}
\newcommand{\eett}{\ensuremath{e^+ e^- \to \tau^+ \tau^-}}
\newcommand{\eemmH}{\ensuremath{e^+ e^- \to \mu^+ \mu^- H}}

\newcommand{\pn}{\ensuremath{\pi^\pm \nu}}
\newcommand{\rn}{\ensuremath{\rho \nu}}
\newcommand{\an}{\ensuremath{(a_1 \to 3\pi^\pm) \nu}}

\newcommand{\ptm}{\ensuremath{p_{T}^{\textrm{\scriptsize \em miss}}}}

\begin{document}

\begin{frontmatter}

\title{ 
Tau 
lepton reconstruction at collider experiments using impact parameters
}

\author{Daniel Jeans}
\address{Department of Physics,
Graduate School of Science,
The University of Tokyo.}

\begin{abstract}

We present a method for the reconstruction of events containing hadronically 
decaying $\tau$ leptons at collider experiments. 
This method relies on accurate knowledge of the $\tau$ production
vertex and precise measurement of its decay products. 
The method makes no assumptions about the $\tau$ kinematics,
and is insensitive to momentum loss along the beam direction.
We demonstrate the method using \eemmtt\ events fully simulated in the
ILD detector.
\end{abstract}


\end{frontmatter}


\section{Introduction}

Studies of final states including $\tau$ leptons are of interest at current and future
collider experiments; as an example, the dominant leptonic decay of the Higgs boson is
to a $\tau$ pair.
A new generation of high energy particle colliders \cite{ILC_TDR_summary, CLIC_CDR, CEPC, FCC}
is presently under study. A key scientific aim of these facilities is to 
measure the Higgs boson's properties with great precision, 
important aspects of which involve measurements of the $\tau$ final state. 
An example is the use of measurements of the $\tau$ spin state to probe the CP nature of the Higgs boson.
The detectors being designed for use at these accelerators will be equipped with
vertex detectors providing unprecedented impact parameter resolution (see {\em e.g.}~\cite{ILC_TDR_detectors}), 
giving rise to intriguing possibilities in the
reconstruction
of relatively long-lived states such as the $\tau$ lepton.

We report on a method which uses a high--precision vertex detector together with other 
tracking and calorimetric detectors to fully reconstruct the kinematics of events 
containing hadronically decaying $\tau$s ({\em i.e.} decays in which only one $\nu$ is produced) 
in an unbiased way. 
We outline previously used techniques for the kinematic reconstruction of single-$\nu$ $\tau$ decays in section~\ref{sec:trad}.
In section~\ref{sec:new} we define a new procedure which, in certain topologies, can fully reconstruct the
$\tau$ kinematics with significantly less stringent assumptions than previous approaches. This new method is then
applied to \eemmtt\ events in section~\ref{sec:example}, and we conclude in section~\ref{sec:conc}.

\section{Previous approaches to $\tau$ pair reconstruction}
\label{sec:trad}

In the case of events containing a pair of $\tau$ leptons each decaying to a single neutrino,
the following method, 
which assumes knowledge of the rest-frame and invariant mass of the $\tau$ pair, but no knowledge about the $\tau$ production vertex,
is often used at lepton colliders ({\em e.g.} \cite{aleph_taupol, belle_tau}). 
The $\tau$-pair rest-frame can be assumed to be the centre-of-mass of the colliding beams 
(in the case of the \eett\ process), 
or the frame recoiling against particles produced in conjunction
with the $\tau$ pair, as in the case \eemmtt.
The $\tau$ decay products are then boosted into the assumed $\tau$ pair rest frame, in which the energy of the $\tau$s is defined
by the assumed invariant mass of the $\tau$ pair. 
The $\tau$ mass then constrains the $\tau$ momentum to be at a fixed angle to the momentum of its hadronic
decay products, defining a cone around the hadronic momentum. 
The two cones in an event, one per $\tau$, have either 0, 1, or 2 intersections, corresponding to 
the possible solutions for the $\tau$ momenta.

At hadron colliders, the unknown net momentum along the beam direction results in less available information to constrain
the event kinematics. The invariant mass of $\tau$ pairs can be partially estimated using the invariant masses of visible decay
products and the missing transverse energy, or by applying the approximation that the $\nu$ from $\tau$ decay is collinear with
the visible $\tau$ decay products~\cite{tautau-collin}. 
Another approach is to combined the measured momenta of visible $\tau$ decay products with constraints on the $\tau$ mass and global
event transverse momentum balance, resulting in an under-constrained system. 
The likelihood of the various $\tau$
decay topologies allowed by the constraints can then be used to choose a best solution, or alternatively to associate
a weight to each solution {\em e.g.}~\cite{tautau-pasha, atlas-tautau, cms-tautau}.

If the $\tau$ production vertex is precisely known, the use of the impact parameters of the 
charged $\tau$ daughters (``prongs'') brings additional information. 
The knowledge of the production vertex can come from the reconstruction of particles recoiling against
the $\tau$s, or from {\em a priori} knowledge of the interaction point, if the size of the interaction region is sufficiently smaller
than the impact parameters of the $\tau$ decay products.
The use of the impact parameter vectors of charged $\tau$ daughters, without full reconstruction of $\tau$ decay kinematics, 
in the analysis of Higgs boson CP properties has been described in {\em e.g.} \cite{desch, berge}, 
while their use in fully reconstructing di-$\tau$ systems of 
known invariant mass and momentum have been demonstrated in \cite{rouge, reinhard}.

\section{Method}
\label{sec:new}
In this section we present methods that can, under certain conditions, fully reconstruct a $\tau$ without 
assuming that it belongs to a $\tau$ pair of particular invariant mass or centre-of-mass frame. 
%
%
%
In section \ref{sec:single} we consider the reconstruction of hadronic single prong final states, 
in which a single charged hadron is produced with zero or more neutral hadrons and a single neutrino.
Such decays account for $49.5\%$ of $\tau$ decays.
Multiprong hadronic $\tau$ decays, in which 
three or more charged hadrons, zero or more neutral hadrons, and a single neutrino are produced, account for $15.3\%$ of $\tau$ decays,
are discussed in section \ref{sec:multi}.
%
%
Leptonic decays of the $\tau$ ($35.2\%$) provide significantly less measurable information about its decay kinematics
due to the production of two neutrinos, and are not further considered in this paper. 


The method relies on precise knowledge of the $\tau$ production vertex and the charged prong trajectories,
and on the reconstruction of any neutral hadrons produced in the decay.
Constraints on the invariant mass and lifetime of each $\tau$, 
and on the overall transverse momentum in the event, are then used to determine the $\tau$ momenta.

\subsection{Tau production vertex}

The uncertainty on the $\tau$ production position should be small in comparison to the decay length of the $\tau$
and the typical impact parameters of its decay products.
%
%
In final states in which the $\tau$s are produced together with more than one prompt charged particle, 
the production vertex can be directly reconstructed on an event-by-event basis using the tracks of these particles
(e.g. the $\mu$s in the process \eemmtt). The proposed linear electron-positron colliders \cite{ILC_TDR_summary, CLIC_CDR} 
have rather small 
interaction regions, which may be used as an additional constraint on the interaction point, although this is not done for the 
results presented in this paper.

\subsection{Single prong $\tau$ decays}
\label{sec:single}

\subsubsection{Tau decay plane}

In the case of single prong hadronic $\tau$ decays, the trajectory
of the charged prong (helical in the usual case of a uniform magnetic field)
can be used to define a plane (hereafter called the ``track plane'') which contains two vectors:
${\bf d}$, the vector between the reconstructed interaction point (IP, assumed to be the $\tau$ production vertex) and 
the point on the trajectory closest to the IP (point of closest approach PCA); and
${\bf p}$, the tangent to the trajectory at the PCA.
In the case of 
linear trajectories of the $\tau$ and of the prong between the PCA and the $\tau$ decay vertex, 
the $\tau$ momentum, and therefore also the sum of the momenta of the other decay products of the
$\tau$ (neutrinos and neutral hadrons), are constrained to lie within this plane.

The difference between the reconstructed track plane and the true $\tau$ decay plane
(defined by the $\tau$ and prong momenta)
depends on the decay length of the $\tau$,
the accuracy with which the IP position is known, 
the precision of the charged prong trajectory, 
and the extent to which the linear approximation of the $\tau$ and prong trajectories near the IP is valid\footnote{
The error in this linear approximation scales as the
ratio of the decay length of the $\tau$ to the radius of curvature of the prong: for a prong with $p_T = 10$~GeV/c produced by
a 50~GeV $\tau$ of average lifetime, in a field of 3.5~T, this ratio is $<10^{-3}$.
An iterative approach, in which a first iteration uses the helix parameters at the PCA to the IP, while later iterations
use the helix parameters at the calculated $\tau$ decay position, should reduce any sensitivity to the prong's curvature.}.

\subsubsection{Parameterisation of neutrino momentum}

In this section we describe the parameterisation of the unmeasured neutrino momentum ${\bf q}$, based on the 
measured prong trajectory and neutral hadron momentum.

Neutral particles are measured as clusters in the calorimeters, 
or as identified conversions of photons into $e^+ e^-$ pairs within the tracker volume.
The momentum to be associated to calorimeter clusters can be estimated by assigning the energy of the calorimeter cluster,
a mass hypothesis ({\em e.g.} zero in the case of photon-like clusters, the $K_L$ mass for hadronic clusters), 
and the direction of a straight line connecting the IP and the energy-weighted mean position of the calorimeter cluster\footnote{
Alternative definitions are possible: for example the line connecting a first estimate of the $\tau$ decay position to the identified start of
the calorimetric shower.}.


The three-momentum ${\bf k}$ of the neutral hadronic system can be decomposed into components perpendicular to and within the
track plane: ${\bf k_\perp}$ and ${\bf k_\parallel}$ respectively.
Since the $\tau$ momentum lies within the track plane, 
the hadronic momentum perpendicular to the track plane must be balanced by the neutrino, 
so the perpendicular component of the neutrino momentum ${\bf q_\perp} = -{\bf k_\perp}$.

%
The component of the neutrino momentum within the track plane can completely generally be parameterised as 
\begin{equation}
{\bf q_\parallel} = Q \cdot (  \cos \psi \cdot {\bf \hat{h}_\parallel} + \sin \psi \cdot {\bf \hat{f}} ),~\footnote{We define ${\bf \hat{x}}$ to be a unit vector parallel to ${\bf x}$.}
\end{equation}
where $Q$ is the unknown magnitude of the in-plane component of the neutrino momentum,
${\bf h_\parallel}$ is the component of the total hadronic momentum (${\bf h} = {\bf p} + {\bf k}$) in the track plane, 
and the unit vector 
${\bf \hat{f}} \equiv {\bf f}/|{\bf f}|$, where
${\bf f} = {\bf h_\parallel} \times ( {\bf d} \times {\bf h_\parallel})$, 
is within the plane and perpendicular to ${\bf h_\parallel}$.
%

\begin{figure}
\begin{center}
\includegraphics[width=70mm]{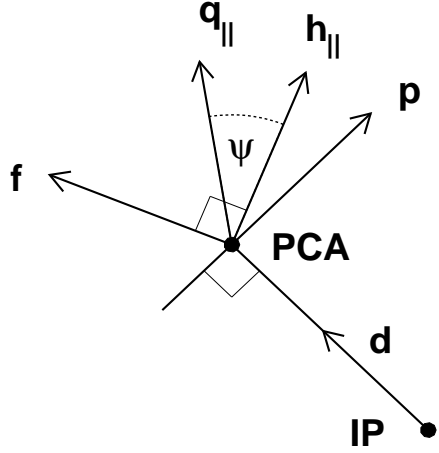}
\caption{
Parameterisation of the neutrino momentum in the track plane (${\bf q_\parallel}$) in terms of the vectors
${\bf h_\parallel}$ and ${\bf f}$, and angle $\psi$. The track plane is defined by the vectors ${\bf p}$ and ${\bf d}$.
}
\label{fig:Qcartoon}
\end{center}
\end{figure}

Four-vectors $p$, $k$, and $q$ corresponding to the three-momenta ${\bf p}$, ${\bf k}$, and ${\bf q}$, can be defined by means
of appropriate invariant mass assumptions.
The invariant mass of the sum of four-vectors $p$, $k$, and $q$ must be equal to the $\tau$
lepton mass $m_{\tau}$. 
This constraint allows us to write an equation involving $Q$ and ${\bf \hat q}_\parallel$ (which in turn depends on $\psi$):
\begin{eqnarray}
m_\tau^2 & = & ( q + h )^2 \nonumber \\ 
        & = & ( \sqrt{ Q^2 + {\bf k}_\perp^2 } + E_h )^2 - (  Q {\bf \hat q}_\parallel - {\bf k}_\perp  + {\bf h})^2
\end{eqnarray}
where $E_h$ is the energy of the hadronic system. This can used to solve for $Q$:
\begin{equation}
Q = \frac{1}{2a} ( -b \pm \sqrt{b^2 - 4ac} ),
\end{equation}
where
\begin{equation}
a = B^2 - C^2 , \ \ b = 2AB , \ \  c = A^2 - C^2 {\bf k_\perp}^2 ,
\end{equation}
and
\begin{equation}
A = m_\tau^2 - m_h^2 - 2 {\bf h}\cdot{\bf k_\perp}, \ \ 
B = 2 {\bf \hat{q}_\parallel} \cdot ({\bf h} - {\bf k_\perp} ), \ \ 
C = 2 E_h,
\end{equation}
where $m_h$ is the invariant mass of the hadronic system.


There are in general two solutions of $Q$ for each choice of $\psi$, which are complex in unphysical regions.
Such complex solutions are rejected.
In the case of two real solutions, we choose to denote the one with a higher energy neutrino in the laboratory frame
as the ``first'' solution, and the other as the ``second''.
The first solution at $\psi = \alpha$ corresponds to the second solution at $\psi = \pi + \alpha$, and {\em vice versa}.
Since the angle between the neutrino and hadrons is typically small in the laboratory (due to the large boost of the $\tau$),
it is convenient to consider the solutions separately, each in the range $-\pi/2 < \psi < \pi/2$.

For each real solution of $Q$, the corresponding $\tau$ momentum can be calculated. In conjunction with the
prong trajectory, this allows the decay length and proper decay time of the $\tau$ to be calculated. In the simple case of $k=0$ 
({\em i.e.} no neutral hadrons in the $\tau$ decay), one solution
for $Q$ corresponds to a negative $\tau$ decay length (the intersection of the $\tau$ trajectory with that of the prong
is on the ``wrong'' side of the IP), and can therefore be discarded.
%
More generally, a likelihood $\lambda$ that the extracted lifetime is consistent with that expected of the $\tau$ can be expressed in terms of
the measured decay length of the $\tau$ in the laboratory $L$,
its Lorentz boost ($\beta, \gamma$), and the mean lifetime of the $\tau$ ($\sim 87 \mu m/c$):
$\lambda = \exp{(-L/(\beta \cdot \gamma \cdot 87 \mu m))}$ if $L>0$, and $\lambda = 0$
otherwise.\footnote{A more sophisticated treatment would take measurement uncertainties into account, 
therefore allowing small negative decay length solutions.}

\subsubsection{Choice of $\psi$}

To determine the value of $\psi$, additional information is required.
One possible approach would be to follow a statistical approach, applying a weight to each possible solution 
based on the likelihood that its decay time is consistent with that expected of the well-known $\tau$ mean lifetime,
and/or that the reconstructed $\tau$ decay kinematics follow the expected distributions.
If a hypothesis is made as to the energy of the $\tau$, or on the invariant mass of a pair of $\tau$s, 
this can also help choose appropriate $\psi$ solutions.

An alternative approach, followed in this paper, is to consider the environment in which the $\tau$ has been produced.
We consider the class of events in which one or more single-$\nu$ decaying $\tau$ leptons have been produced in conjunction with
zero or more well-measured particles, together with zero or more particles escaping along the beam-line (e.g. ISR photons).
Examples of such processes at an electron-positron collider are
two-fermion production \eett, 
Higgs-strahlung $e^+ e^- \to H Z \to (\tau^+ \tau^-) (\mu^+ \mu^-)$, and its major irreducible background
$e^+ e^- \to Z Z \to (\tau^+ \tau^-) (\mu^+ \mu^-)$.
In such events, the overall $p_T$ of the $\tau$s and other visible particles is balanced, while there may be
non-zero net momentum along the beam-line due to ISR. 
A natural way in which to determine the $\psi$ angles in an event (one per $\tau$ decay) is to choose
that combination which minimises the 
magnitude of the missing transverse momentum (\ptm) in
the event. This method has the advantage of making no assumptions about
$\tau$ kinematics or about undetected particles escaping along the beam-line 
-- also indispensable for application at hadron collider experiments, but does require that no additional neutrinos
are produced in the same event.
In the case of several solutions for which \ptm\  is consistent with zero,
the reconstructed $\tau$ lifetimes can be used to determine the most likely one, or each solution could be
weighted by a likelihood based on its lifetime and/or $\tau$ decay kinematics.


\subsection{Multiprong $\tau$ decays}
\label{sec:multi}

In the case of multiprong $\tau$ decays, the prongs can be fitted to a common vertex, whose position ${\bf V}$ corresponds to the
point at which the $\tau$ lepton decayed. Together with a well-known production position ${\bf P}$, this gives an estimate of the 
direction of the $\tau$ momentum, in the approximation that the $\tau$ trajectory is linear\footnote{
We note that the $\tau$ momentum direction can be directly measured for any decay mode 
if the $\tau$ decay length is sufficiently large for it to produce hits in the vertex detector.
}.
The $\tau$ momentum ${\bf p_\tau}$ can be written as ${\bf p_\tau} = t {\bf \hat{r}}$,
where $t$ is the magnitude of the $\tau$ momentum, 
and 
${\bf r} = {\bf V} - {\bf P}$.

The neutrino momentum ${\bf q}$ can then be written as ${\bf q} = {\bf p_\tau} - {\bf h}$, where ${\bf h}$ is the measured momentum
of the visible hadronic system. Defining appropriate four-vectors $q$ and $h$ for the neutrino and hadronic system respectively, and requiring that the
invariant mass of the decay products is equal to the $\tau$ mass $m_\tau$, gives
\begin{eqnarray}
m_\tau^2 & = & ( q + h )^2 \nonumber \\ 
         & = & ( |{\bf q}| + E_h )^2 - ({\bf q} + {\bf h})^2 \nonumber \\
         & = & ( | t {\bf \hat{r}}  - {\bf h}|  + E_h )^2 - t^2,
\end{eqnarray}
where $E_h$ is the energy of the hadronic system. This can be solved for $t$, giving
\begin{equation}
t = \frac{1}{2a} ( b \pm \sqrt{ b^2 - 4ac } ),
\end{equation}
where
\begin{equation}
a = E_h^2 - ({\bf \hat{r}} \cdot {\bf h})^2, \ \ 
b = ({\bf \hat{r}} \cdot {\bf h}) (m_\tau^2 + m_h^2), \ \ 
c = - \frac{ (m_\tau^2 + m_h^2)^2 }{4} + m_\tau^2 E_h^2 .
\end{equation}
There are in general two possible solutions for $t$ which satisfy the $m_\tau$ constraint, 
however they are not guaranteed to be real.
The full tau decay kinematics can be calculated for each real solution, including the decay length and
proper decay time of the $\tau$. 
In appropriate event topologies, the \ptm\  of the event can be used to choose between these two solutions,
otherwise the previously introduced lifetime likelihood could be used.

Since the $\tau$ leptons produced at high energy colliders are typically highly boosted, the opening angle of the 
multiprong jet is usually small. As a consequence, the precision with which the $\tau$ decay vertex is reconstructed
is significantly worse along direction of the $\tau$ jet than along the perpendicular directions. In the example considered 
later in this paper, in which $\tau$s are produced in Higgs boson decays, 
the length of the major axis of the vertex position error ellipsoid is typically of order $100 \mu m$, 
while the other two axes have lengths of order $2 \mu m$.

In this case, it turns out to be better to consider the major axis of the vertex error ellipsoid as a single 
trajectory, and to analyse the event using the procedure developed for single prong decays in the previous section.
In this way, the $\tau$ decay plane is defined by the IP and the centre and major axis of the 
vertex ellipsoid, and the exact decay position along the major axis direction is left as a free parameter.
The distance between the fitted and reconstructed vertex positions,
normalised by the uncertainty on the reconstructed vertex position, can be used to define a vertex likelihood,
which can be used when choosing between several possible solutions.

\section{Example application}
\label{sec:example}

We apply this method to \eemmH\ events, in which the Higgs boson decays to a pair of $\tau$s.
Such events are usually selected by considering the mass recoiling against the di-$\mu$ system, which has a peak 
at the Higgs boson mass. However the distribution of this recoil mass has a long tail due to beamstrahlung and ISR,
particularly at higher centre-of-mass energies well above threshold. If the invariant mass of the $\tau \tau$ system can be directly
reconstructed, a rather cleaner selection of such events should be possible. The full reconstruction of the $\tau$ decay kinematics and
the $\tau \tau$ centre-of-mass frame also allows the best use of the spin information of the $\tau$s: the properties of the 
$\tau$ decays can be used to define their {\em polarimeter vectors}, and the correlations between the $\tau$ polarimeters 
can be used to measure the CP-nature of the Higgs boson ({\em e.g.} \cite{rouge}).

Events were generated at a centre-of-mass energy of 250~GeV using {\sc whizard} v2.2.2\cite{whizard}, 
assuming a Higgs boson mass of 125~GeV, and including effects due to
initial state radiation and beamstrahlung (by {\sc circe1}).
The $\tau$s were decayed by {\sc tauola++} v1.1.4\cite{TAUOLA}. 
Three scenarios were considered: both $\tau$s decaying to \pn; 
both decaying to $\pi^\pm \pi^0 \nu_\tau$; and both decaying to \an. 
The simplest hadronic $\tau$ decay mode is \pn, 
but accounts for only 11.5\% of $\tau$ decays,
while the 
\rn \ 
mode has the largest branching ratio of 26.0\%.\footnote{
The vast majority of $\tau \to \pi^\pm \pi^0 \nu_\tau$ decays
proceed via the $\rho^\pm$, so we use \rn\ 
as shorthand for $\pi^\pm \pi^0 \nu_\tau$ in this paper.}

The resulting events we passed through the {\sc geant4}-based {\sc mokka} simulation of the ILD\_o1\_v05 detector model~\cite{ILC_TDR_detectors}.
This model consists of a vertex detector with three double-layers of silicon pixel detectors, 
a silicon strip-based inner tracker and forward tracking disks,
a large time projection chamber within a silicon tracking envelope, followed by highly granular calorimeters: a silicon-tungsten ECAL
and scintillator-iron HCAL. These are placed within a solenoid producing a 3.5T magnetic field, surrounded by an instrumented iron
flux return yoke.
The simulated energy deposits within the active detectors were passed through the standard ILD digitisation procedures to 
simulate detector signals.

\subsection{Event reconstruction}

\begin{figure}
\begin{center}
\includegraphics[width=0.49\textwidth]{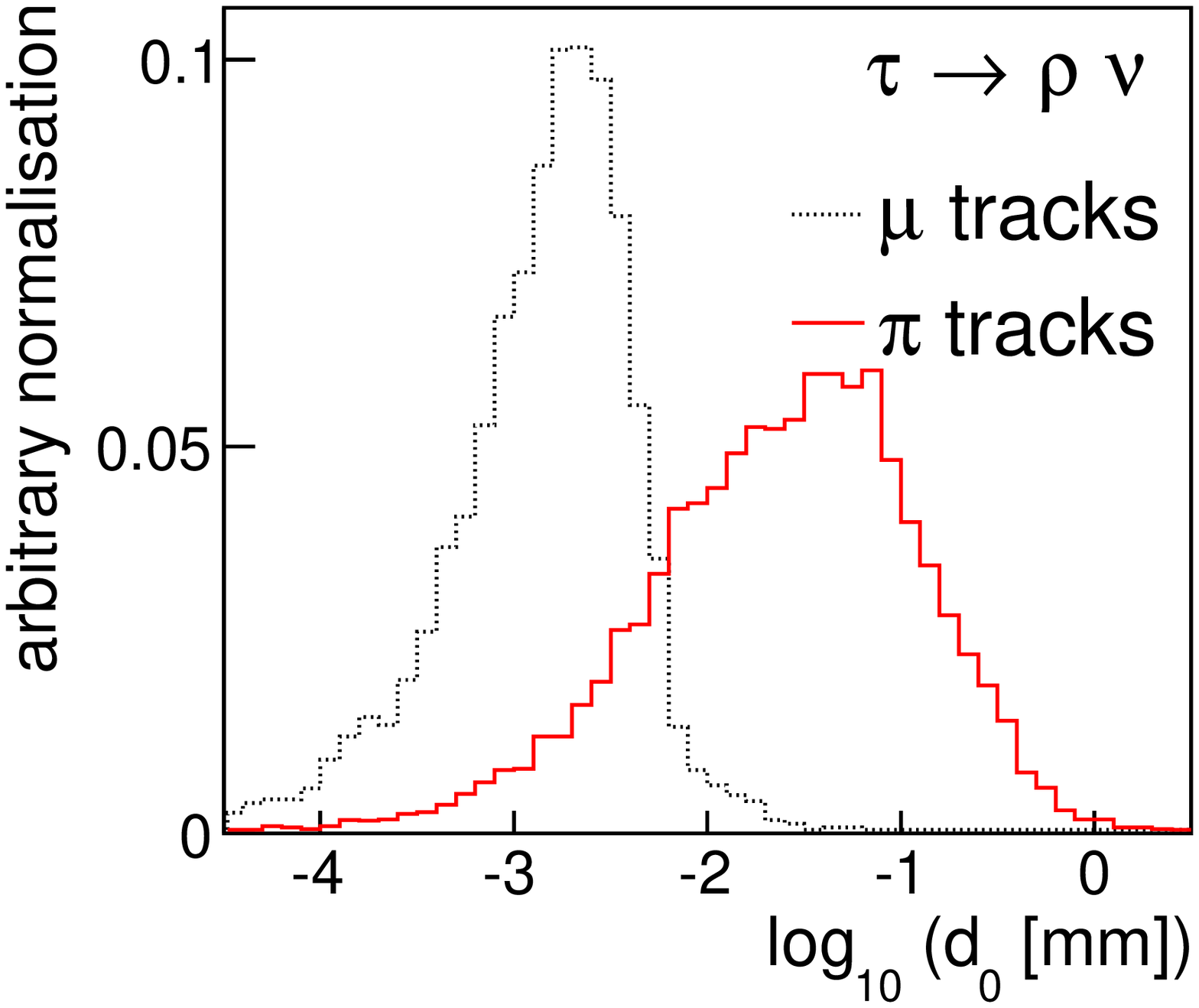} 
\includegraphics[width=0.49\textwidth]{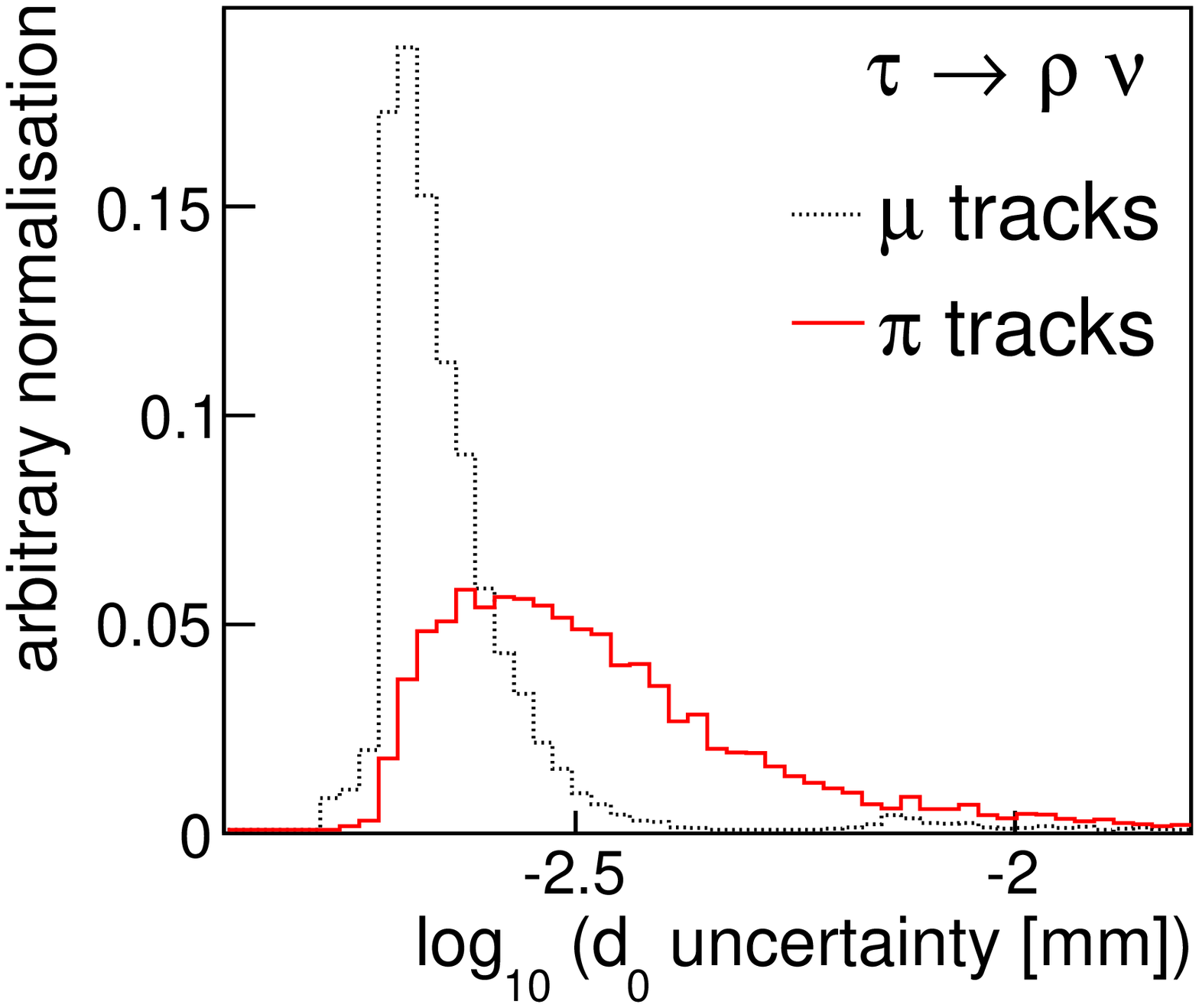} \\
\includegraphics[width=0.49\textwidth]{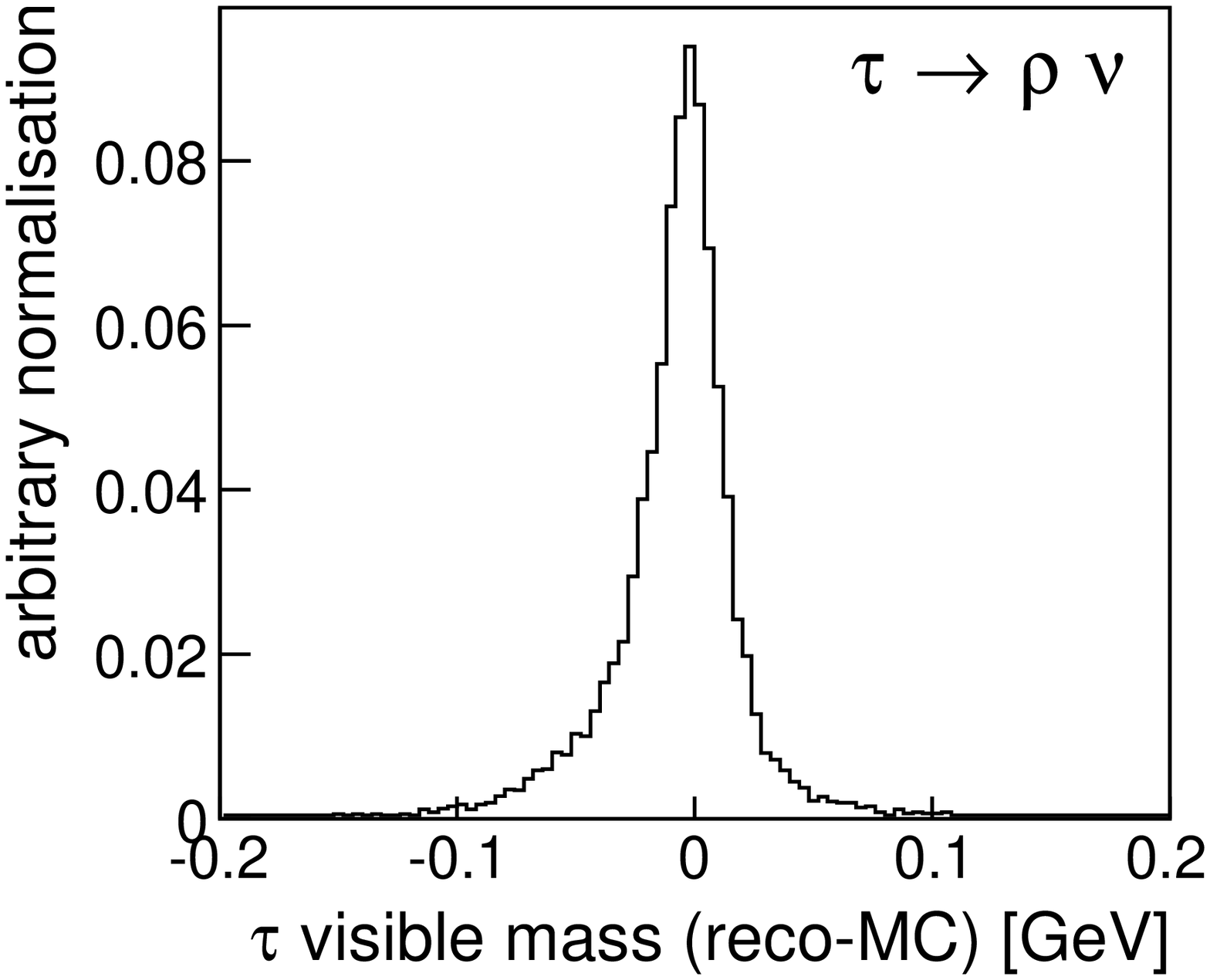}
\includegraphics[width=0.49\textwidth]{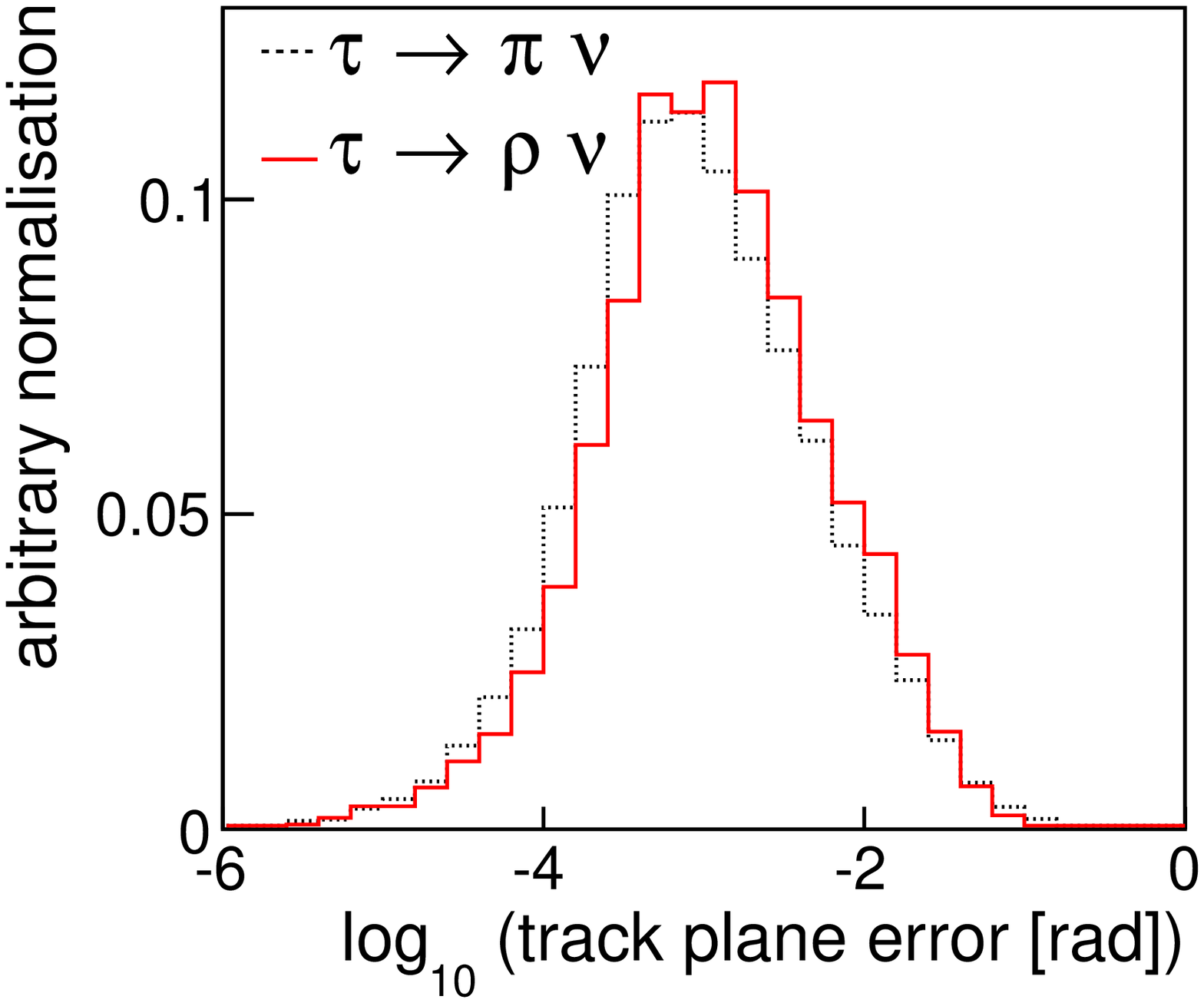}
\caption{
Properties of the considered \eemmH\ events.
Top: the measured  2-d impact parameter $d_0$ [left] and its uncertainty [right], for prompt $\mu$ tracks, and $\pi$ tracks
from $\tau \rightarrow \rho \nu$ decay.
Bottom left: the difference between the reconstructed and true visible $\tau$ mass in $\tau \rightarrow \rho \nu$ decays.
Bottom right: the angle between the reconstructed track plane and the true direction of the $\tau$ momentum.
}
\label{fig:reco}
\end{center}
\end{figure}

The standard ILD reconstruction software was used to reconstruct charged particle tracks and calorimeter clusters.
For the results shown in this paper, the reconstructed tracks were associated to primary muons and charged pions from $\tau$ decays
on the basis of matching to the simulated particle directions.
In a full analysis, muon and pion identification should be performed. This is rather simple task in detectors
with highly granular calorimeters such as ILD, which should not present any great difficulties.
Distributions of the impact parameter in the plane perpendicular to the beam $d_0$, and of its uncertainty,
are shown in fig.~\ref{fig:reco} for both prompt $\mu$ tracks and $\pi$ tracks from $\tau$ decay.
The two $\mu$ tracks were fitted to a common vertex to measure the event IP using the {\sc lcfivertex}\cite{lcfivertex} package,
achieving a typical precision on the IP position of better than $3 \mu m$ in all three dimensions.

Clusters identified in the electromagnetic calorimeter by the 
{\sc garlic}\cite{GARLIC} and {\sc pandorapfa}\cite{PandoraPFA} algorithms were treated as photon
candidates.
Each cluster was assigned a momentum with magnitude equal to the reconstructed cluster energy and direction parallel to a line
joining the nominal IP and the energy-weighted mean position of the cluster.
Clusters were associated to $\pi^0$s and their parent $\tau$s based on matching the cluster momentum direction to the simulated photon directions.
Events were rejected if not all the simulated $\mu$, $\pi$ and photons could be matched to reconstructed particles.
Such cases are typically due to
imperfect reconstruction algorithms or the
interactions of particles in the tracker volume.

In order to improve the effective photon energy resolution,
clusters associated to a $\pi^0$ were subjected to a constrained 
kinematic fit imposing the $\pi^0$ mass by varying their energies. 
No special effort was made to correct invariant mass biases occurring when reconstructing two overlapping clusters from $\pi^0$ decay.
Figure~\ref{fig:reco} shows the precision with which the 
visible invariant mass is reconstructed in $\tau \to \rho \nu$ decays.
The accuracy with which the reconstructed track plane approximates the true $\tau$ decay plane is demonstrated in the bottom right plot of 
fig.~\ref{fig:reco}, in which the ``track plane error'' is defined as the angle that the true $\tau$ momentum makes to the
reconstructed track plane.

The reconstructed tracks from multiprong $\tau$ decays were fitted to a common vertex, again using the 
{\sc lcfivertex} package. In three-prong decays via the $a_1$, the ellipsoid describing the vertex position 
uncertainty typically has a major axis length of $100\mu m$, and the other two axes around $2 \mu m$.

\subsection{Dependence of \ptm\  and the lifetime likelihood on $\psi^\pm$}

\begin{figure}
\begin{center}
\includegraphics[width=0.49\textwidth]{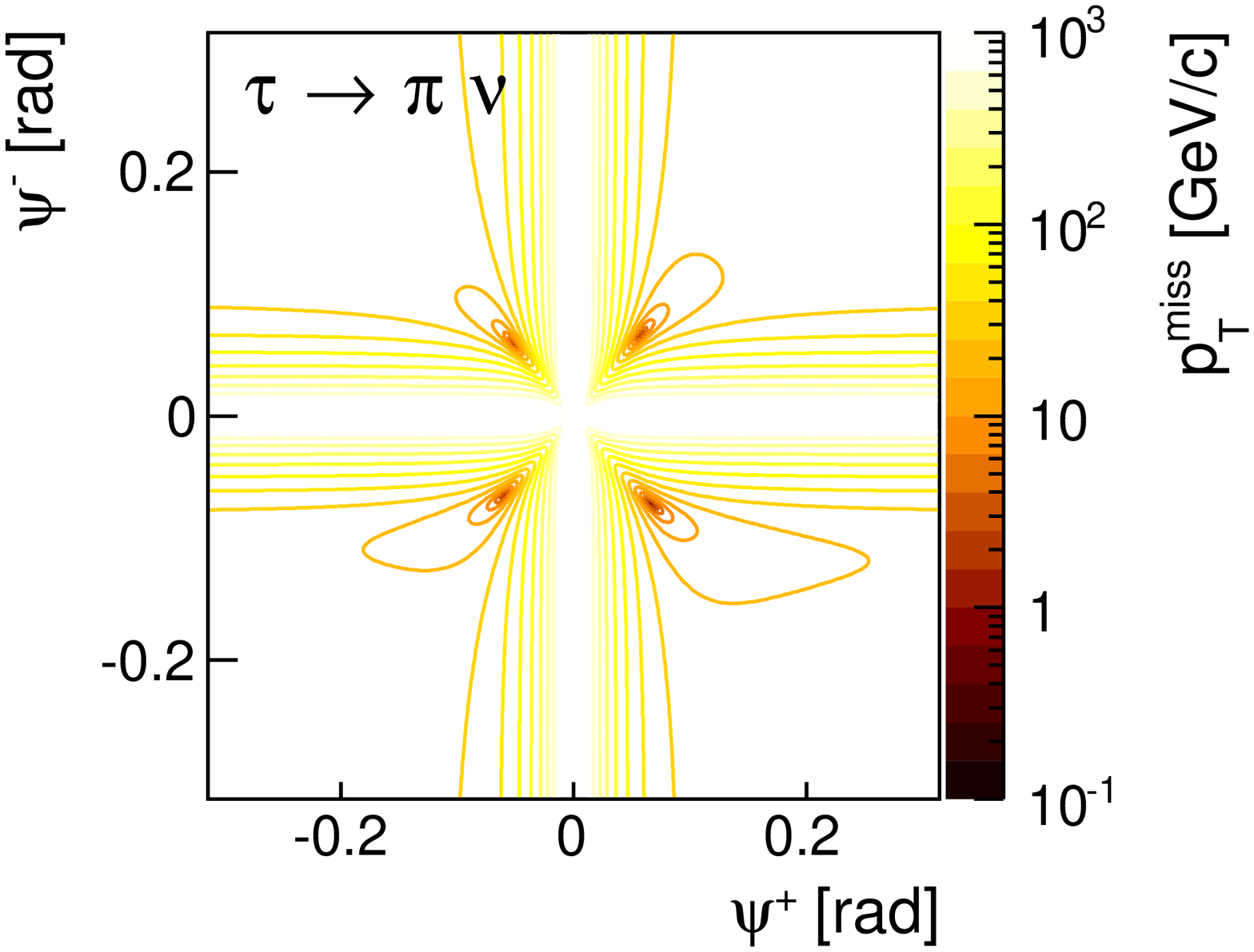}
\includegraphics[width=0.49\textwidth]{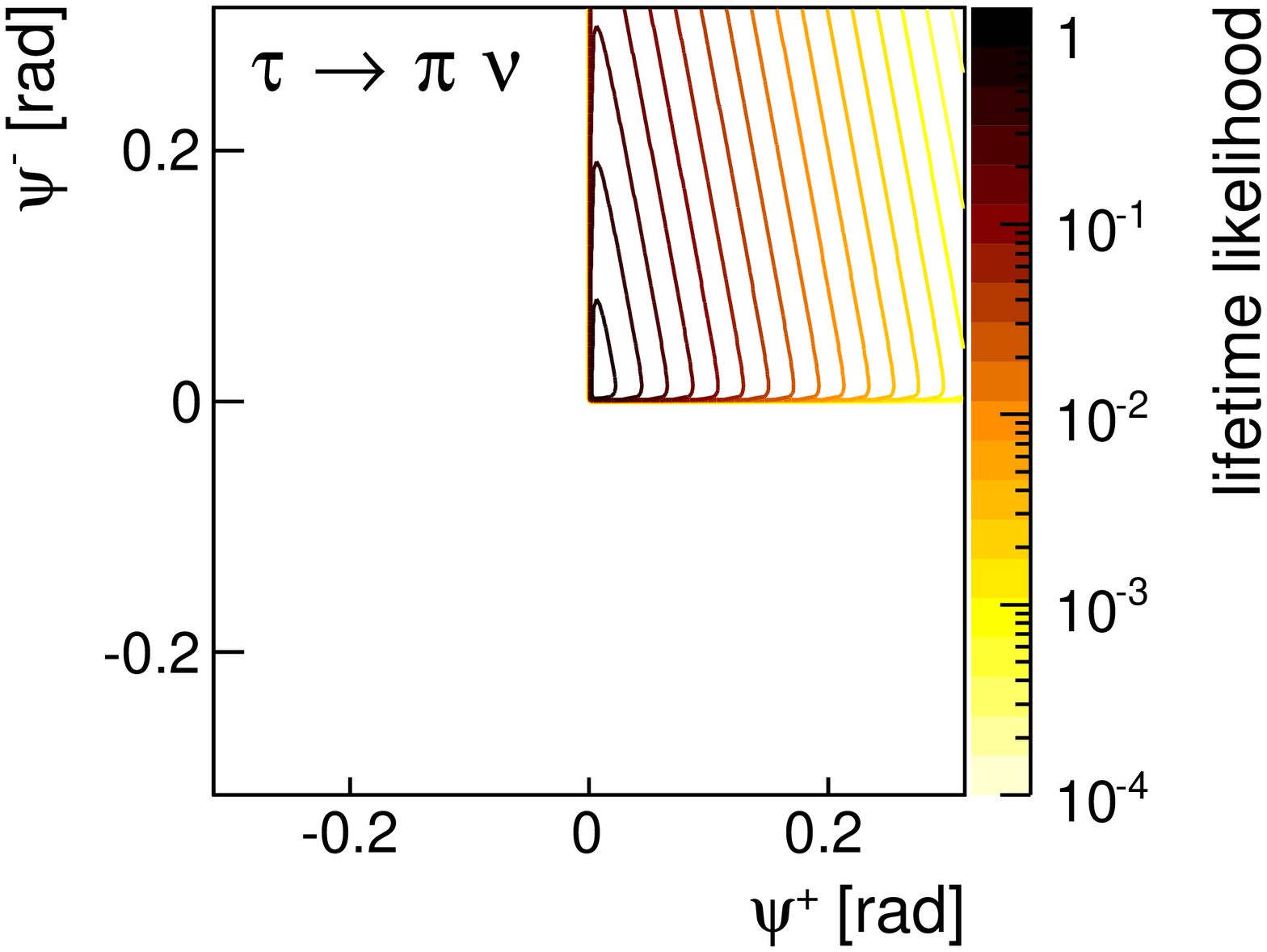} \\
\includegraphics[width=0.49\textwidth]{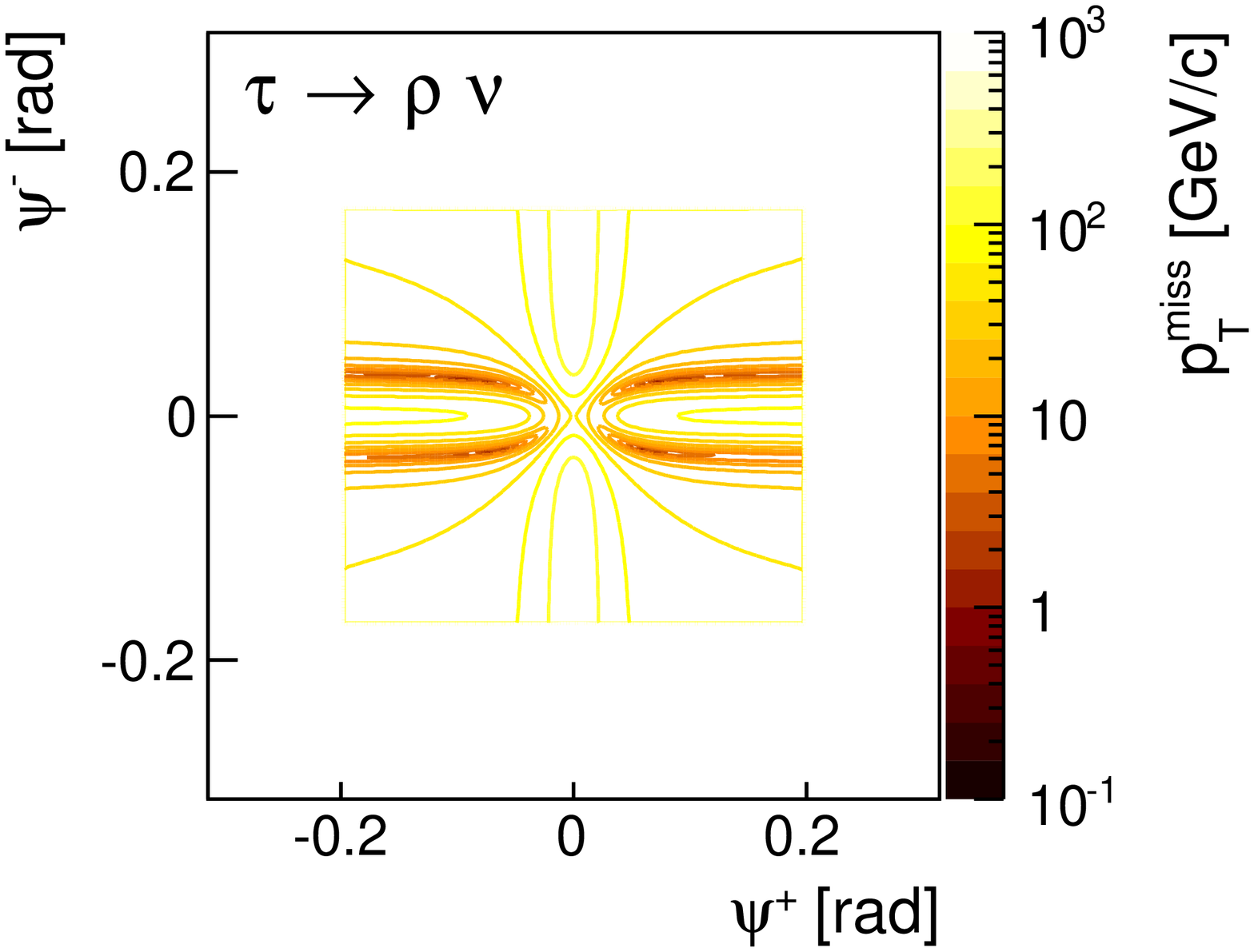}
\includegraphics[width=0.49\textwidth]{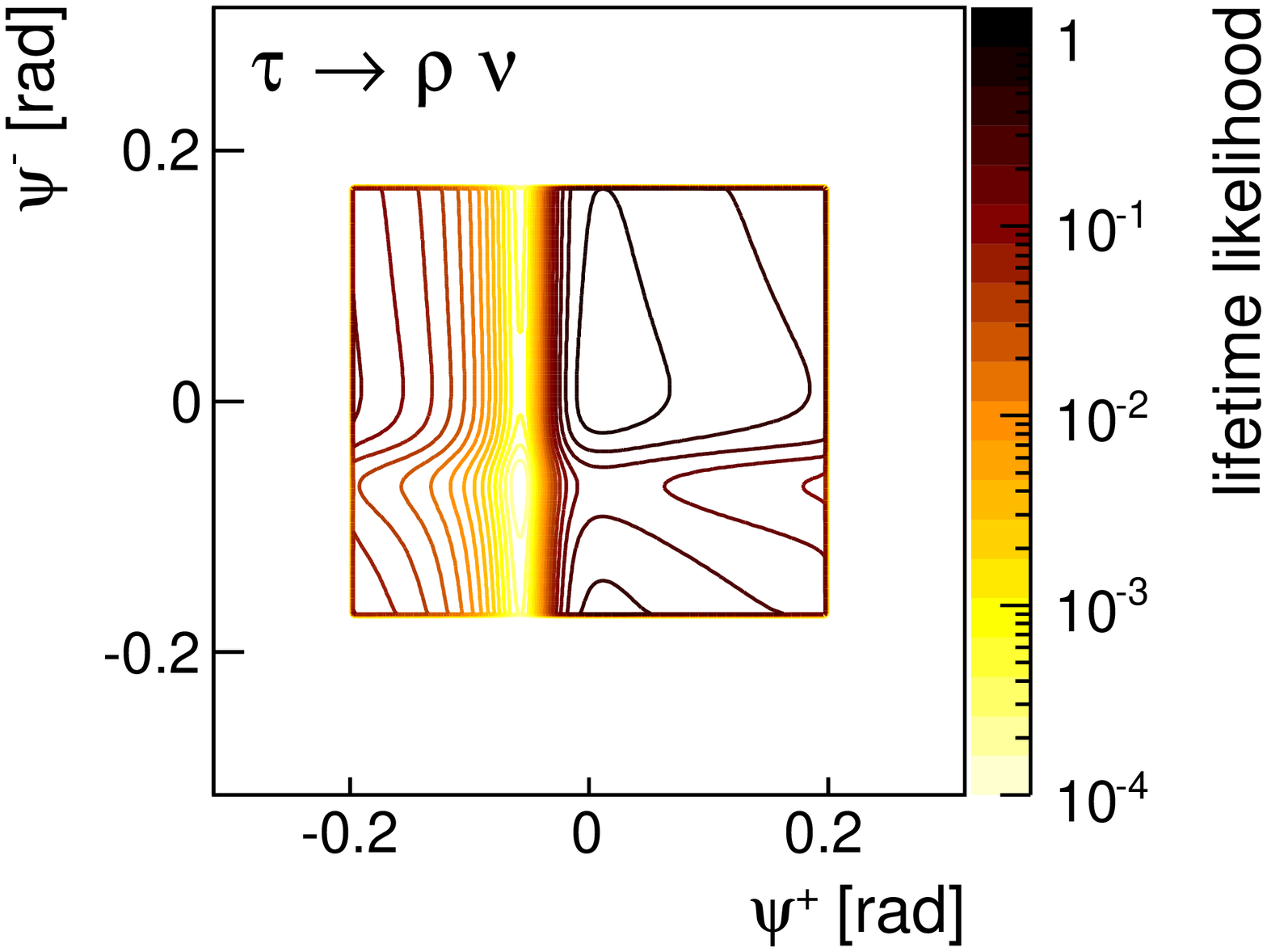} \\
\caption{Example of two \eemmtt\ events in which 
both $\tau \rightarrow \pn$\ [top], or both $\tau \rightarrow \rn$\ [bottom]. 
Contours of \ptm  [left] and lifetime likelihood [right] as a function of the two angles $\psi^\pm$ are shown.
Regions with darker contours (lower \ptm, higher lifetime likelihood) are preferred. 
In the upper event no positive decay length solutions were found for $\psi^\pm < 0$,
while in the lower event, no physical solutions were found for $|\psi^\pm| > \sim 0.2$.
}

\label{fig:pipievent}
\end{center}
\end{figure}

Figure~\ref{fig:pipievent} shows the dependence of the event \ptm\  (defined as the magnitude of the 
component of ${\bf P} = \sum_i {\bf p}_i$ transverse to the beam-line, 
where the index $i$ runs over the momenta ${\bf p}_i$ of the $\tau$s and other reconstructed particles in the event) 
and lifetime likelihood on the 
two $\tau$ decay angles $\psi^\pm$ in two events, one in which both $\tau \to \pn$, the other in which both $\tau \to \rn$.
The distributions are shown for the case in which the first Q solution is used for each $\tau$
(which is where the best solution was found in these two events).
Four minima at which $\ptm \sim 0$ are seen, one in each quadrant. 
The lifetime likelihood (shown for the same Q solution),
together with the \ptm\  at the minimum, allows the best minimum to be chosen.

The {\sc minuit} package in the {\sc root}\cite{root} analysis framework was used to find the minima of the \ptm\  distributions 
separately in each of the four quadrants 
bounded by $\psi^\pm = 0, \pm \pi/2$,
in each of the four $Q$-solution combinations.
If no minimum was found, or either $\tau$ had a negative reconstructed decay length at the minimum, the quadrant was rejected.
Of the remaining possible solutions, the one with the smallest value of \ptm\ 
was chosen.

\subsection{Results}

\subsubsection{Single prong decays}

\begin{figure}
\begin{center}
\includegraphics[width=0.49\textwidth]{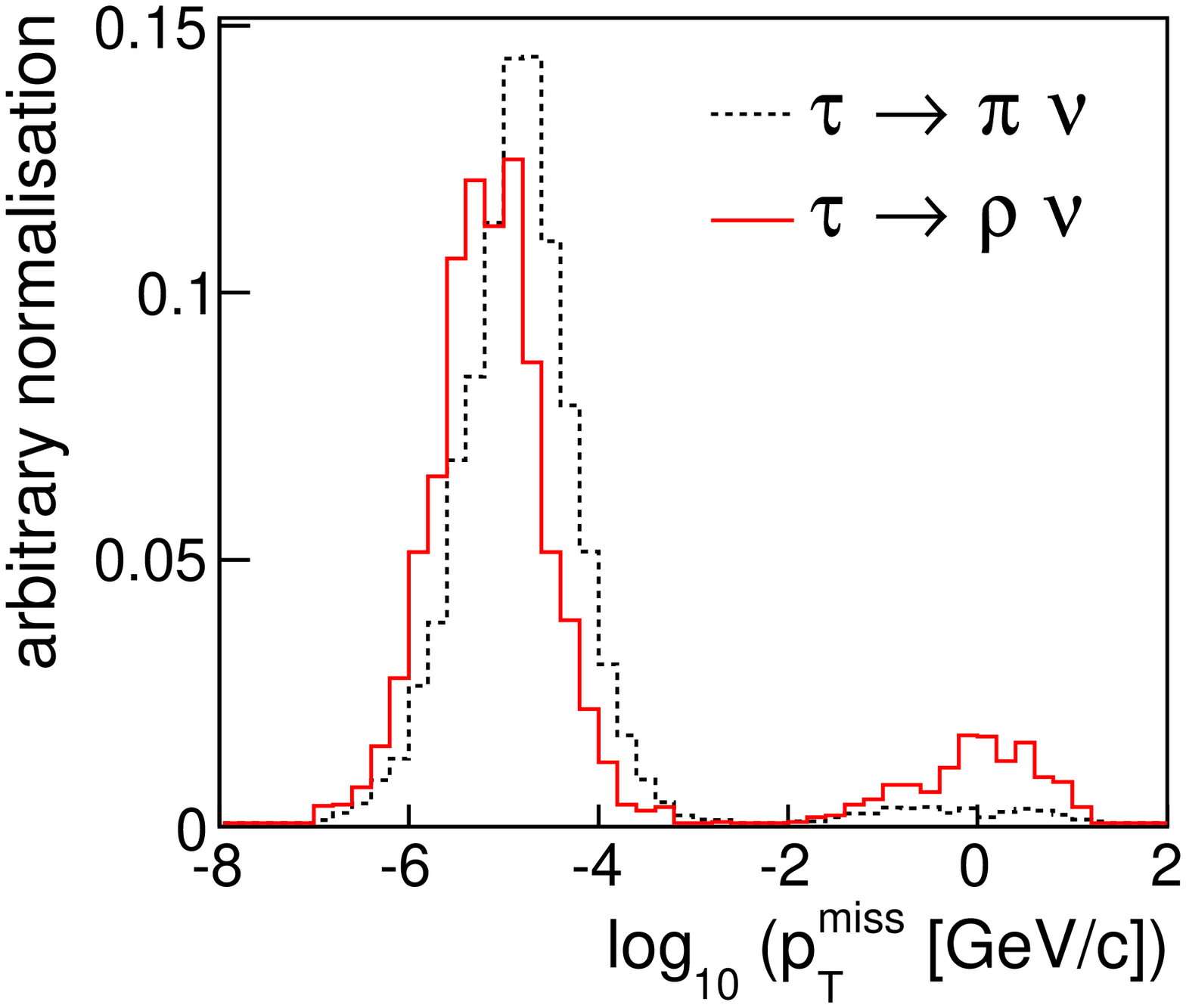}
\includegraphics[width=0.49\textwidth]{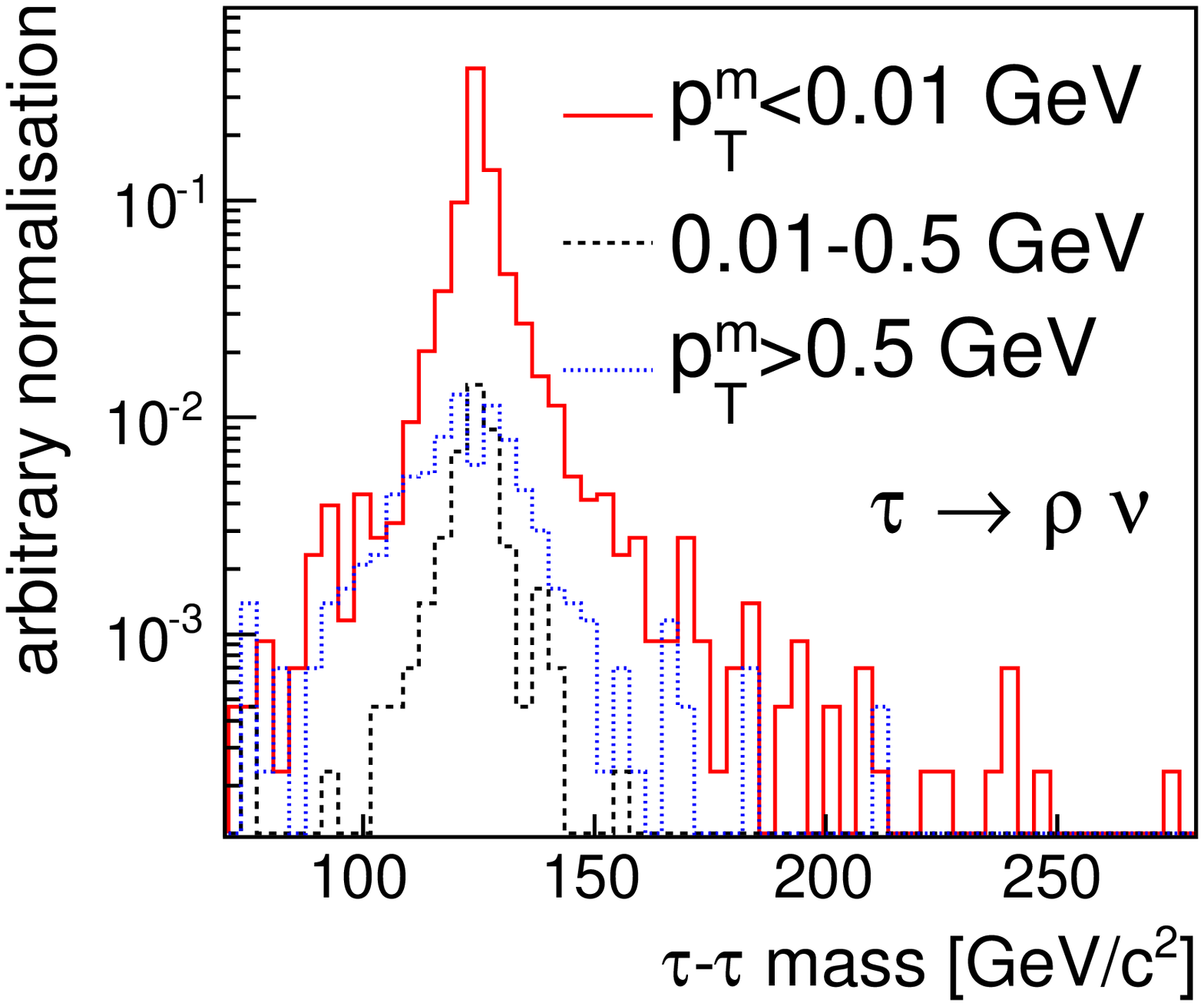} \\
\includegraphics[width=0.49\textwidth]{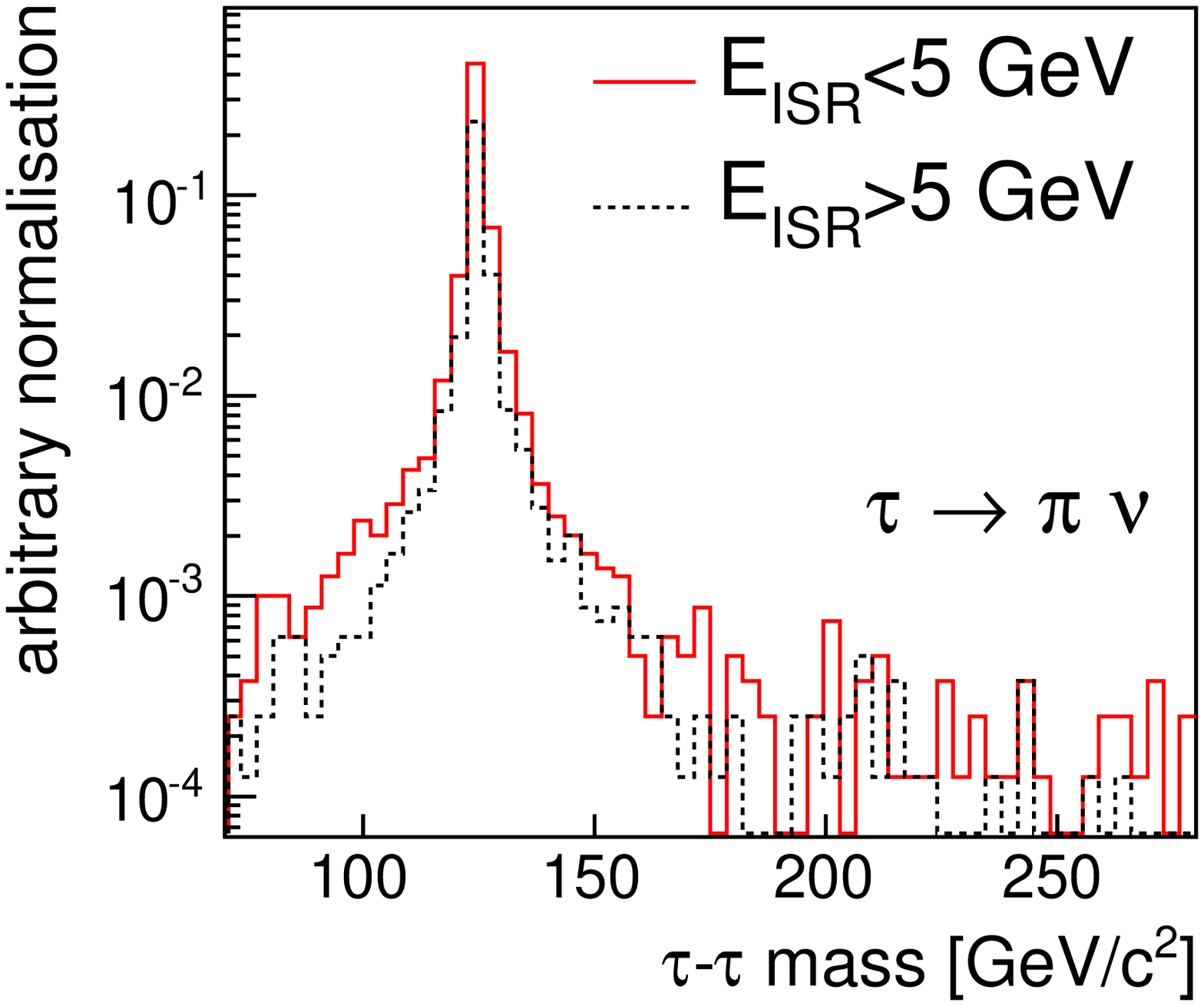}
\includegraphics[width=0.49\textwidth]{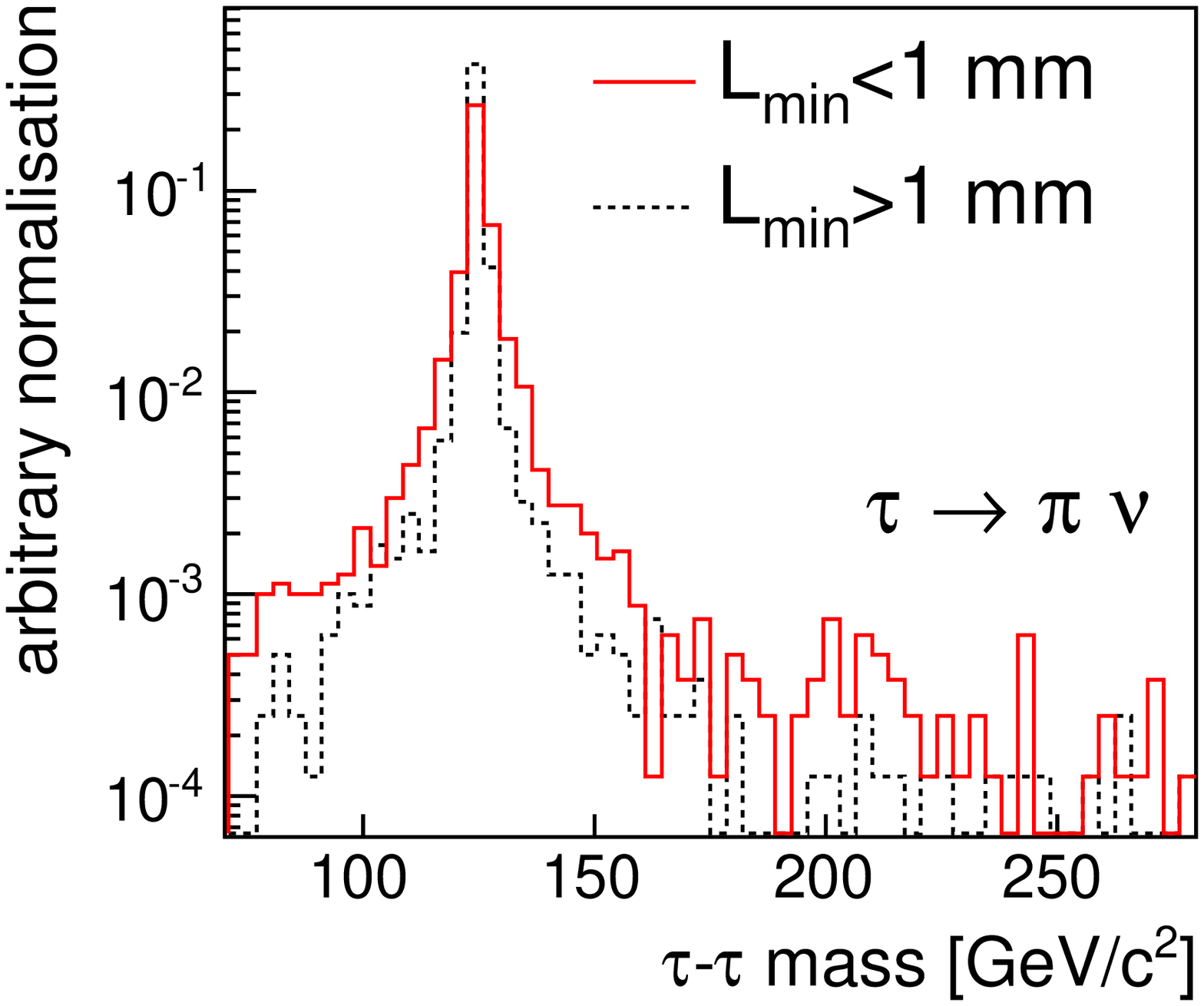} 
\caption{
Properties of the considered \eemmH\ events after full $\tau$ reconstruction.
The value of the event \ptm\  at the chosen solution [top left].
The  $\tau$ pair mass distributions in different ranges of 
$p_{T}^{\textrm{\scriptsize \em m(iss)}}$
[top right], 
the total ISR/beamstrahlung energy $\textrm{E}_\textrm{ISR}$ [bottom left], and
$L_{min}$, the smaller of the two true $\tau$ decay lengths [bottom right].
Except in the top left plot,
the relative normalisations of the various contributions are as in the event samples.
}
\label{fig:pt_mass}
\end{center}
\end{figure}

Of events in which reconstructed objects could be unambiguously matched to simulated particles,
no good solution was found in 0.4\% (2.4\%) of events in the \pn (\rn) channel.
This inefficiency is due to cases in which the reconstructed visible mass is larger than the $\tau$ mass, or in 
which no solution with positive decay length is found.
In fig.~\ref{fig:pt_mass} we show the distribution of the chosen solutions' \ptm, 
for events in which both $\tau$s were forced to decay to either \pn\ or \rn.
A rather good minimum ($\ptm < 1~$MeV) is found in the large majority of cases, but a fraction of events have a 
larger \ptm.
These events with large \ptm\  are due to mis-reconstruction of the event, and make up a larger fraction of
\rn\ 
than 
\pn\ 
decays due to the relatively worse precision of the calorimeters compared to the tracker.
Events in which the reconstructed \ptm\  is larger than around 0.5 GeV/c show a significantly wider peak in the $\tau-\tau$ mass distribution, 
as shown in fig.~\ref{fig:pt_mass}.
The mass distribution in the case of 
\pn\ (\rn)
decays has a central peak with a 
Gaussian width $\sigma \sim 0.6$~GeV (1.1~GeV), and non-Gaussian tails to both higher and lower values. 
Of 
\pn\ (\rn)
events with $\ptm < 0.5$~GeV/c,
74\% (67\%) lie within $3 \sigma$, and 95\% (89\%) within 10~GeV, of the peak position.

\begin{figure}
\begin{center}
\includegraphics[width=0.49\textwidth]{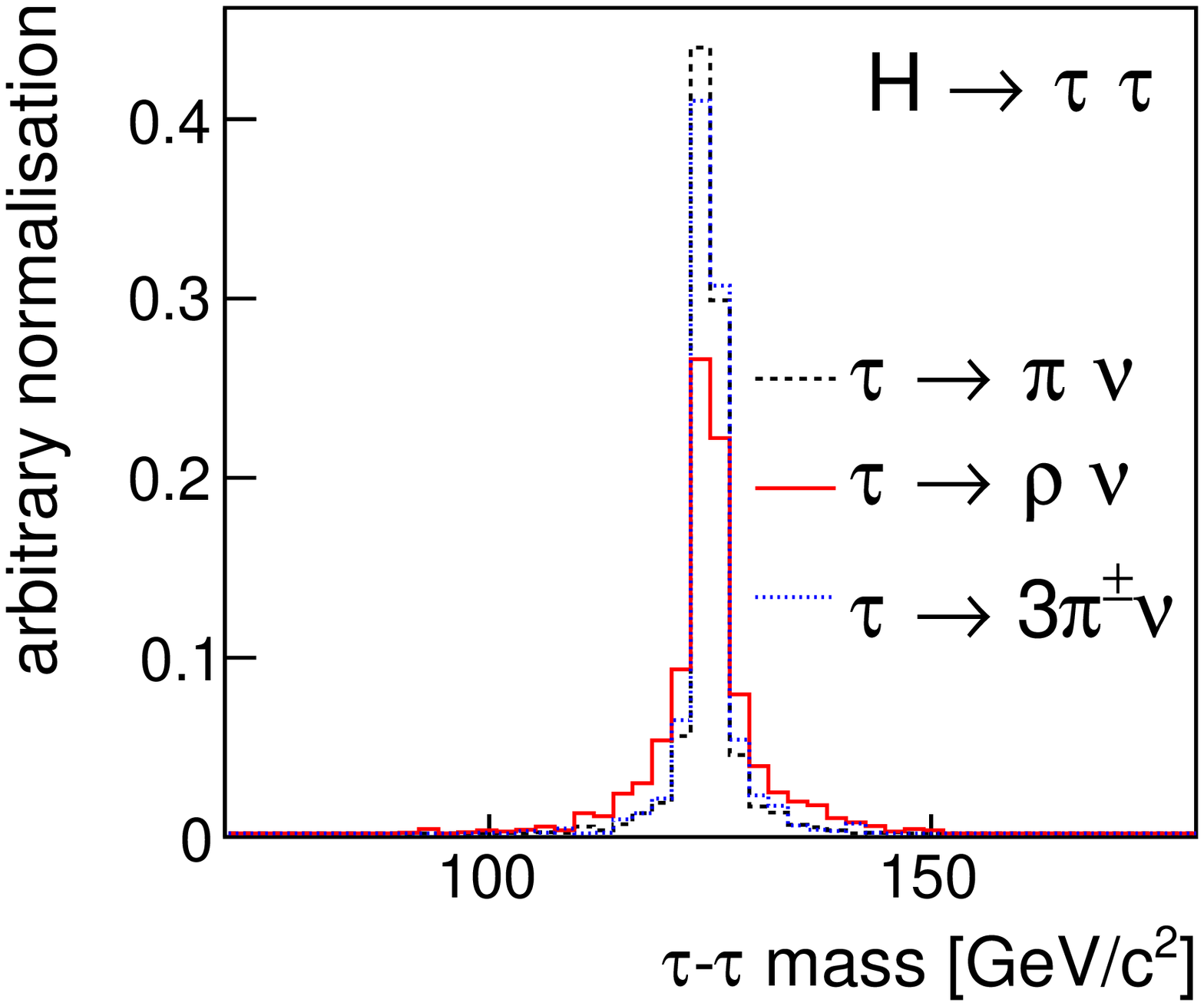}
\includegraphics[width=0.49\textwidth]{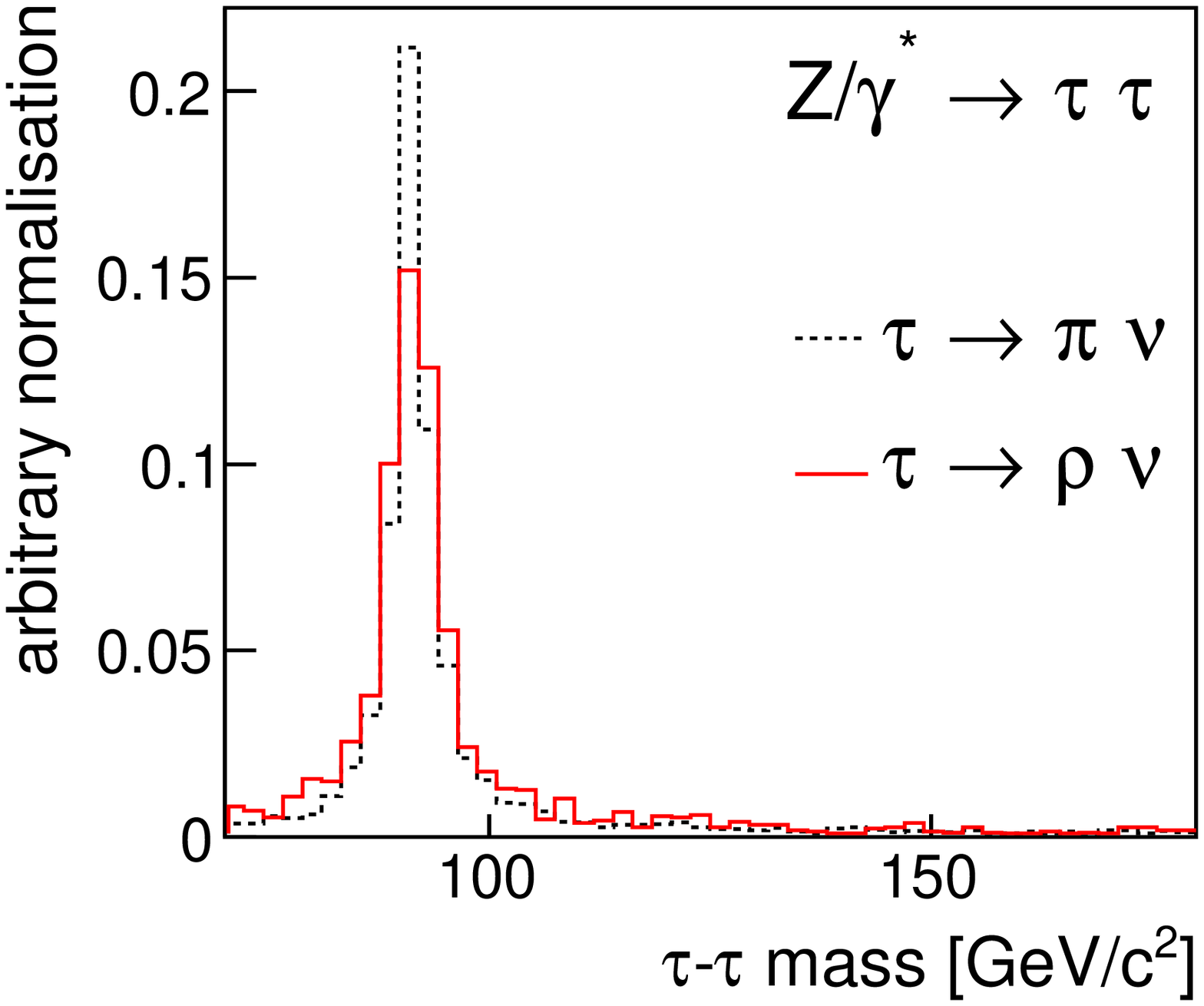} \\
\caption{
The reconstructed mass of the $\tau$ pair in \eemmtt events in which the 
$\tau$ pair was produced via a Higgs (left) or $Z^0/\gamma^*$ (right).
}
\label{fig:massZH}
\end{center}
\end{figure}

The same figure also shows the reconstructed invariant mass of the two $\tau$ system
in events with small or large ISR/beamstrahlung energy (demonstrating that the method is independent of the 
total centre-of-mass energy and boosts along the beam-line), 
and on the smaller of the true $\tau$ decay lengths in the laboratory (showing that events 
with longer decay lengths tend to be better reconstructed).
Figure~\ref{fig:massZH} shows the  $\tau$ pair mass in the \eemmH\ event sample, 
and a second \eemmtt\ sample in which the 
Higgs boson contribution was effectively turned off.
Clear mass peaks due to the Higgs and $Z^0$ bosons can be seen, the widths of which allow a 
rather clean separation between these processes in both the considered $\tau$ decay modes.

\subsubsection{Multiprong decays}

Multiprong $\tau$ decays, studied using the $\tau \to \an$
channel, 
are reconstructed using two methods. In the ``vertex'' method, the reconstructed vertex position
is directly used to constrain the $\tau$ direction, and the $\tau$ mass constraint
used to find two possible solutions for the neutrino momentum. 
No real solution is found in around a quarter of $\tau$ decays.
The event \ptm\  is used to choose the best solution. 

\begin{figure}
\begin{center}
\includegraphics[width=0.49\textwidth]{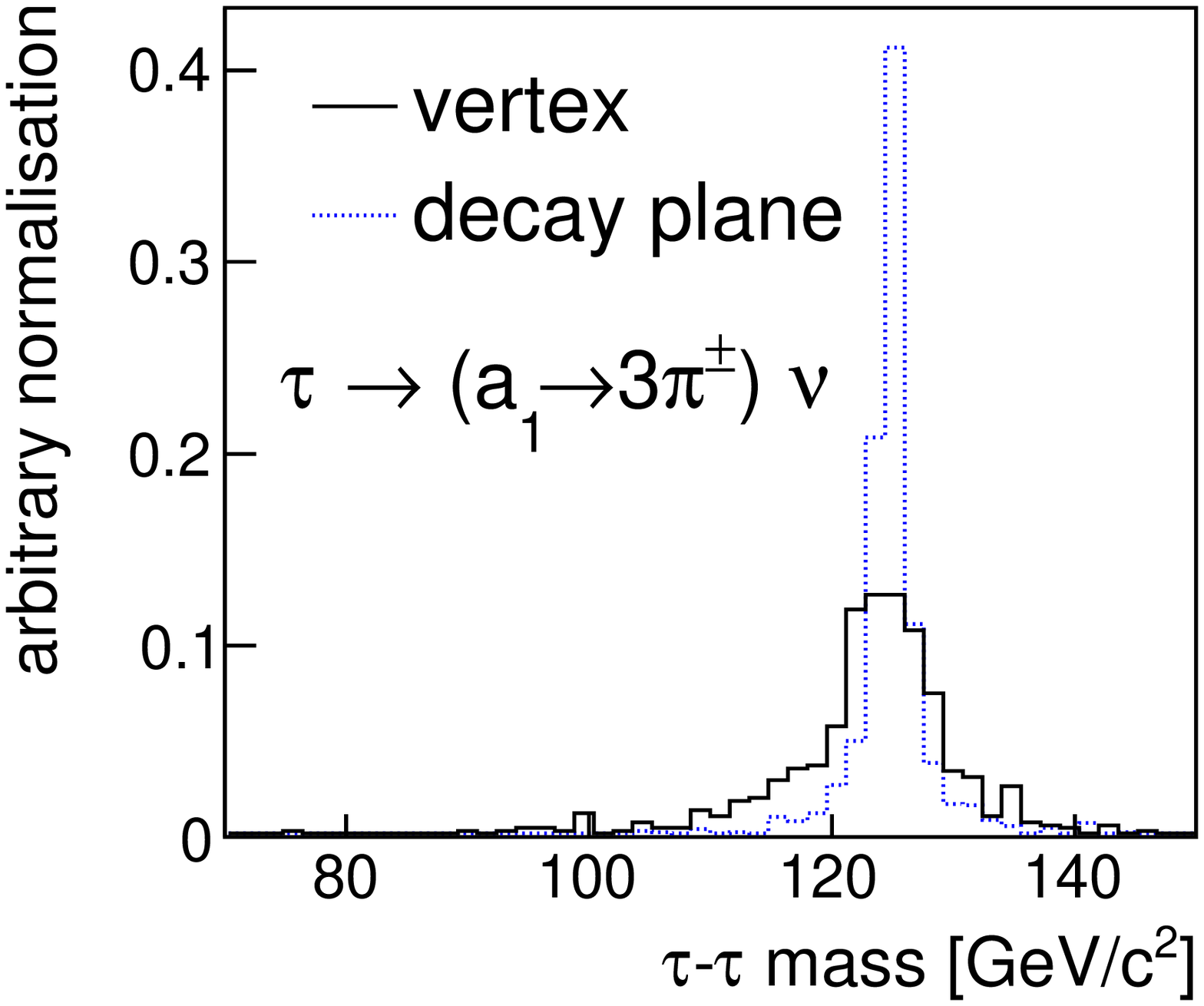}
\caption{
The reconstructed mass of the $\tau$ pair in \eemmtt events in which both 
$\tau$s decay into three charged pions. We compare the results of two $\tau$ reconstruction methods:
directly using the reconstructed vertex position, and using the position and major axis of the 
vertex ellipsoid to define the $\tau$ decay plane.
}
\label{fig:massZHaa}
\end{center}
\end{figure}

In the ``decay plane'' method,
the central position and major axis of the vertex position error ellipsoid are used to define
the $\tau$ decay plane, which is then used in the same way as for single prong decays.
Candidate $\tau$ solutions are required to have fitted and reconstructed $\tau$ decay vertices in
the same hemisphere. 
The event candidate solution with smallest \ptm\  is selected.
When using this method, no good candidate solution was identified in around 4.5\% of events.
%
Figure \ref{fig:massZHaa} shows a comparison of these two multiprong methods.
As well as having a much higher efficiency for identifying a solution,
the ``decay plane'' approach results in significantly better mass resolution.
The di-$\tau$ invariant mass distribution for multiprong $a_1$ decays 
obtained by using the ``decay plane'' method
are compared to other decay channels in fig.~\ref{fig:massZH}, 
showing that multiprong decays are reconstructed with a similar precision 
as $\tau \to \pn$ decays.

\begin{figure}
\begin{center}
\includegraphics[width=0.49\textwidth]{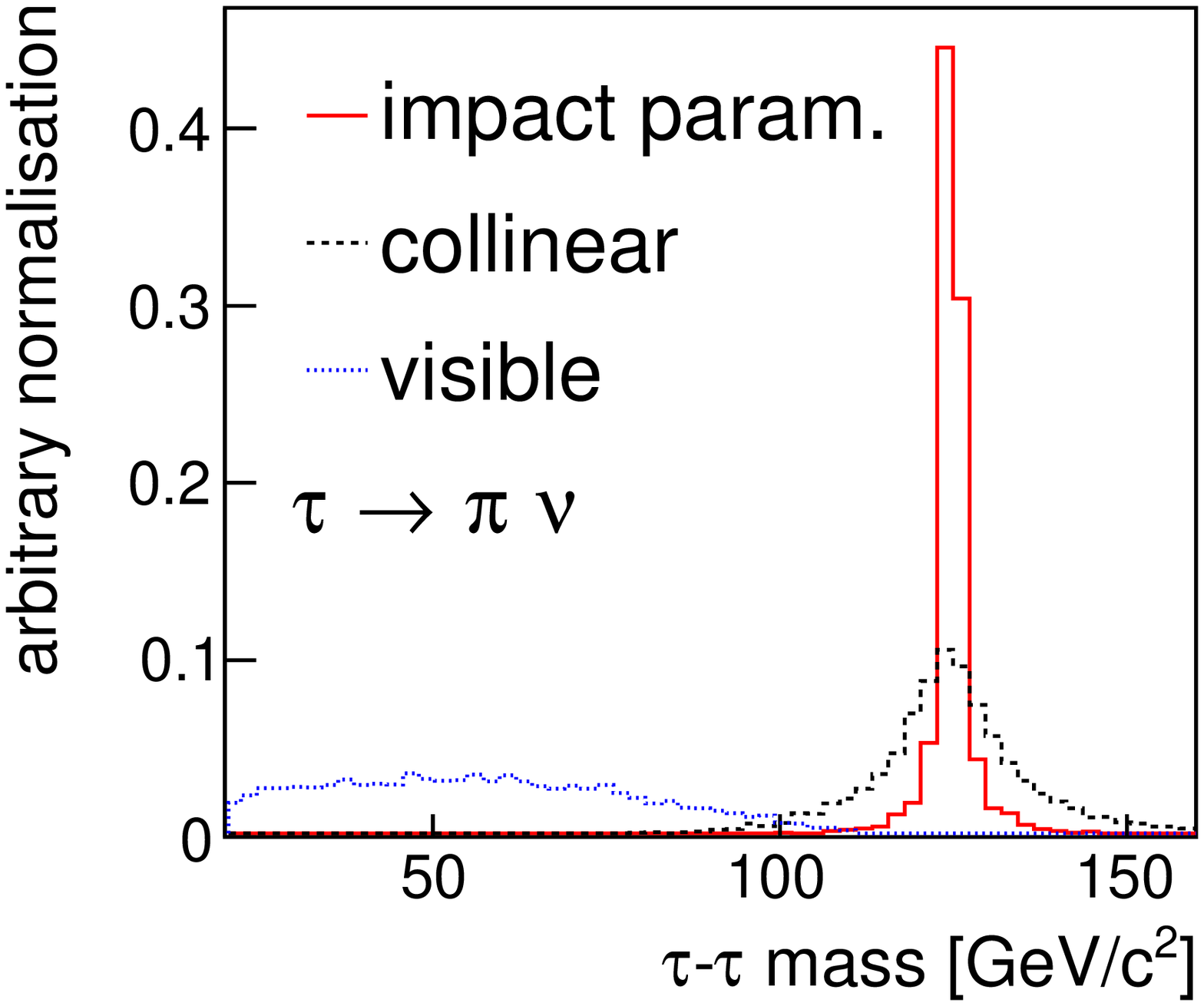}
\includegraphics[width=0.49\textwidth]{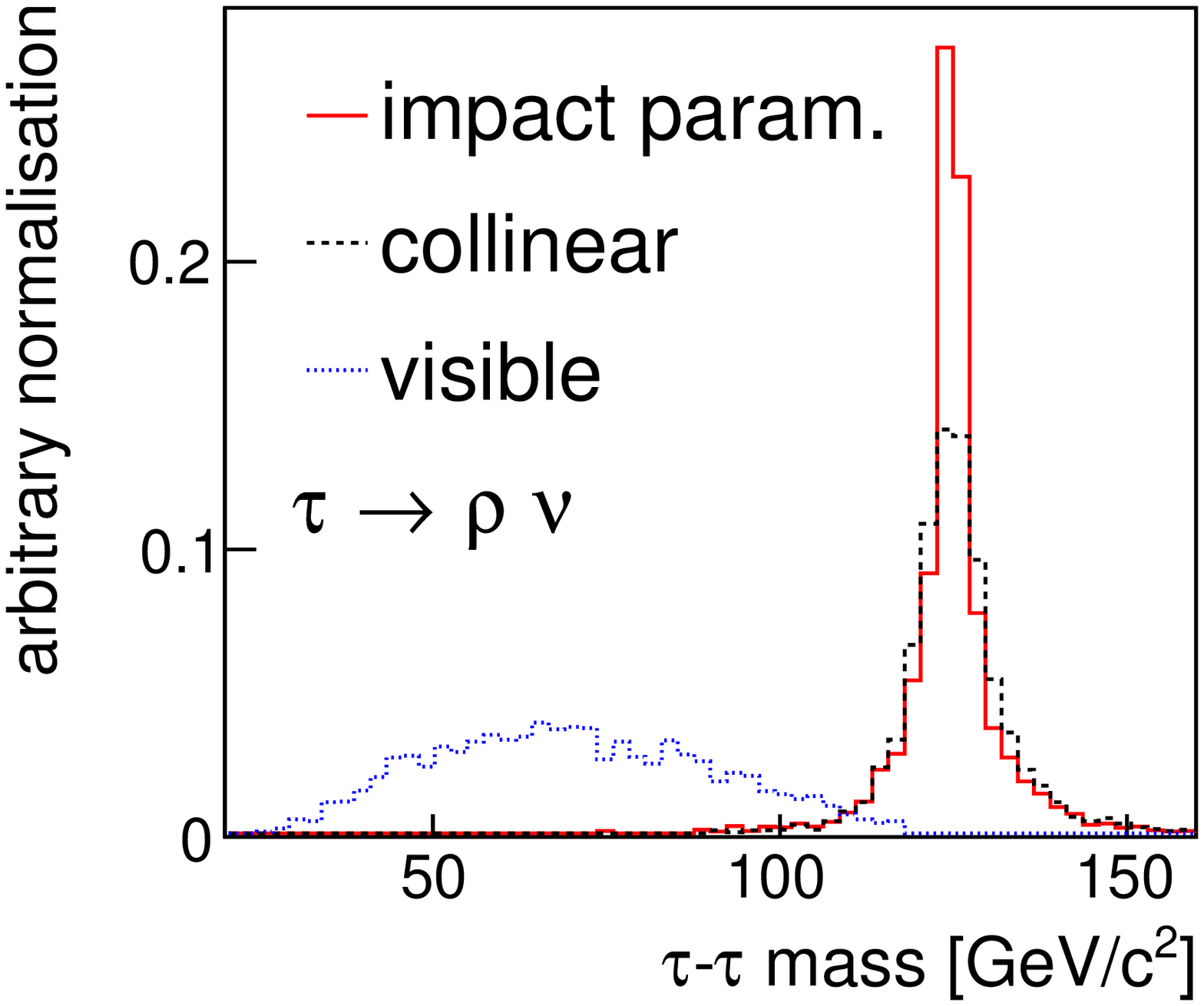} 
\caption{
Comparison of the reconstructed $\tau$ pair invariant mass using different methods:
the impact parameter-based method developed in this paper,
the collinear approximation, and 
only the visible decay products. 
}
\label{fig:collin}
\end{center}
\end{figure}

In fig.\ref{fig:collin} we compare $\tau$ pair mass reconstruction using the methods developed in this paper with alternative approaches.
The simplest is to use the invariant mass of only the visible decay products, therefore ignoring the neutrino contribution.
This clearly underestimates the invariant mass, and results in a rather wide distribution. 
The collinear approximation assumes that the neutrinos are parallel to the visible tau decay products, 
and further requires the event's $p_T$ to be balanced, allowing the neutrino energies to be estimated.
The mass distribution resulting from the collinear approximation is peaked at the correct value,
however it is significantly wider than that achieved by the methods developed in this paper.
The difference in resolution is particularly significant in the \pn\ 
decay channel, where the
impact parameter method gives a very sharp distribution.

\section{Conclusion}
\label{sec:conc}

An approach to the reconstruction of hadronic $\tau$s has been presented, which can be used in events
in which the $\tau$ production vertex is well known, and the trajectory of its charged decay products is
precisely measured. This method works only in events in which no undetected particles with significant $p_T$
are produced together with the $\tau$s. In contrast to other methods, no assumptions are made on the 
centre-of-mass frame or the invariant mass of a $\tau$-pair system, nor on the $\tau$ energies. 
The method is insensitive to 
momentum balance
along the beam-line,
so this method should work on {\em e.g.} \eemmH\ events produced both near, and well above, threshold.

The example analysed in this paper, in which a $\tau$ pair recoils against a $\mu$ pair, is the 
case in which the event \ptm\  is most precisely measured. 
The Higgs boson will more commonly recoil against a hadronic system.
These events are more difficult to reconstruct,
and will have a less precisely measured \ptm. The use of a constrained kinematic fit, taking into
account the resolution with which the various quantities are measured, should lead to improved results.

This reconstruction technique can be used at lepton colliders, as demonstrated in this paper, provided the 
precision of the tracking detectors is sufficient, particularly in the estimation of charged particle 
trajectories near the interaction point. 
To be used at hadron collider experiments, the $\tau$ decay products must be sufficiently well identified and reconstructed,
and all particles produced in the same interaction unambiguously identified in order to accurately estimate the \ptm\  
associated to the interaction. 
This latter point suggests that cases in which $\tau$s recoil against a relatively simple system, {\em e.g.} $\mu-\mu$ as 
considered in this paper or a single high ${\textrm p_T}$ jet, 
should be the most promising in the richer environment of hadronic collisions at high luminosity, 
in which multiple interactions typically occur in each event.
The method is insensitive to the unknown net momentum along the beam direction inherent in hadronic collisions.

If applied to Higgs boson decays into $\tau$ leptons, the use of this method will allow clean separation of signal
events from irreducible backgrounds from {\em e.g.} Z decays, and allow precise reconstruction of the $\tau$
spin information, providing a sensitive probe to measure the CP nature of the Higgs boson.

\section*{Acknowledgements}

I thank H.~Videau and S.~Komamiya for helpful comments on the manuscript.
This work was funded by the MEXT KAKENHI Grant-in-Aid for Scientific Research on Innovative Areas, no. 23104007.

%

\bibliographystyle{elsarticle-num} 

\bibliography{taureco}

\end{document}